\documentclass[12pt]{siamart190516}

\usepackage{latexsym}

\usepackage{amssymb}
\usepackage{pifont}
\usepackage{graphicx}
\usepackage{amsfonts}%
\usepackage{amsmath}

\title{perpetual callable American volatility options in a mean-reverting volatility model}

\author{Hsuan-Ku Liu\thanks{Department of Mathematics and Information Education, National Taipei University of Education, Taiwan ({\tt hkliu.nccu@gmail.com}).}}

\begin{document}

\maketitle

\begin{abstract}
This paper investigates problems associated with the valuation of callable American volatility put options. 
Our approach involves modeling volatility dynamics as a mean-reverting 3/2 volatility process. 
We first propose a pricing formula for the perpetual American knock-out put.
Under the given conditions, the value of perpetual callable American volatility put options is discussed.
\end{abstract}



\pagestyle{myheadings}
\thispagestyle{plain}
\markboth{Hsuan-Ku Liu}{The perpetual callable American volatility option}

\section{Introduction}
\subsection{Review of the callable American option}
A standard American option is a contract which gives the holder the right to exercise it prior to the maturity date. When it is exercised early, the holder gains the amounts 
$(K-S_t)^+$, in the put and $(S_t-K)^+$ in the call, where $S_t$ denotes the price of the underlying asset. For the American option, the holder chooses an optimal strategy, say, the early exercise price,
$S_t^*$ to maximize the value at any time prior to the maturity.
 Hence, the fair price of the American options leads to the solution to the optimal stopping problem $V(S,t)=\sup_{\tau}E[e^{-r\tau}\psi(S_\tau)|S_t=S]$ for certain stochastic processes. The solution can be obtained by solving a free boundary problem (McKean, 1965), where the free boundary is regarded as the early exercise price (or the optimal trading strategy for the holder). 
So far, a considerable number of studies have been published to solve the free boundary problem arising from the American option pricing problem (for example, Geske and Johnson, 1984;  Jacka 1992; Kuske and Keller, 1998; Merton, 1973).

A callable American option is an American option embedded with an additional feature that the seller may cancel the contract at any finite time (before maturity). 
At any time before expiry, the buyer can exercise the option early to maximize the option's value while the seller can call it back to minimize the option's value. 
Kifer (2000) first introduced the callable American option and analyzed the pricing problem based on the theory of optimal stopping games (Dynkin, 1969).
 In an illiquid market, it may not be possible for the seller to construct a hedge portfolio to compensate for the short position. Hence, recalling (or cancelling) the option is possibly the better hedging strategy. 
The strategic recalling can be regarded as an efficient instrument to limit risk when the seller expects stock prices to fall. Recalling such contracts might provide an effective way of reducing undesirable positions in turbulent times (Kifer, 2000). The cheapest super-hedging strategy for the seller of an American put option can be a trivial one consisting of an investment of K units in the riskless bank account.

Under the Black-Scholes framework, a considerable number of studies have begun to study how embedding the costly cancellation right of the seller to cancel the contract prior to maturity may affect its value.  Kyprianou (2004) provided an explicit pricing formulas for the perpetual Israeli $\delta$-penalty American put and the perpetual Israeli $\delta$-penalty Russian put. Ekstr\"{o}m (2006) explicitly determined the value of the perpetual game option by using the connection between excessive functions and concave functions. In the finite maturity case, Kunita and Seko (2004) first investigated the pricing problem of $\delta$-penalty game call options and Kuhn and Kyprianou (2007) characterized the value function as a compound exotic option. Ekstr\"{o}m and Villeneuve (2006) provided the price formula of the perpetual $\delta$-penalty game call on stocks in the absence of a dividend payment. Yam, Yung and Zhou (2014) and Emmerling (2012) considered the pricing problem of the $\delta$-penalty game call on a stock with a dividend payment.

\subsection{Volatility options and Mean-reverting processes}
With the rise and fall of stock prices, investors have been looking for financial instruments to reduce the variability of the price of their investment portfolios. To date, the American option has become
a popular instrument for the buyer in relation to the variability of the price. 
For the seller, the callable feature embedded in the American option also reduces the risk of the price fluctuating sharply.
However, according to empirical research, the volatility of asset returns will change over time. In addition to the price risk, the participants (buyer and seller) usually face the risk of volatility in the securities market. Options written on a security are designed primarily to deal with price risk. Positions in these options may not be sufficient to hedge all the uncertainties in volatility. Therefore, volatility derivatives become a natural candidate for hedging volatility risk. In recent decades, much of the investors' interest in volatility options (or volatility derivatives) seems to have been responsible for the collapse of major financial institutions (e.g., Barings Bank and Long-Term Capital Management). This is because the collapses were accompanied by a dramatic increase in volatility. The volatility option is an instrument whose payoff depends explicitly on some volatility indicator. Today, one of the most popular volatility indicators is the VIX, which is the implied volatility of 30-day S\&P 500 options. 

Recently, VIX-related products (ETNs, futures and options) have become popular financial instruments, for both hedging and speculation.
Brenner and Galai (1989) and Whaley (1993) proposed options written on a volatility index in the geometric Brownian motion. Based on empirical evidence, volatility is mean-reverting (for example, French, Schwert, and Stambaugh 1987; Harvey and Whaley 1991). Grünbichler and Longstaff (1996) assumed that the volatility risk premium is proportional to the volatility index and evaluated volatility futures and options based on the mean reversion square root process (MRSRP):
$$
dx_t=(a-bx_t)dt+c\sqrt{x_t}dz,
$$
where $x_t$ denotes the index of the volatility at time $t$.
Several mean-reverting volatility models are now applied to price the volatility derivatives, such as the inverse square-root process (Detemple and Osakwe, 2000), log mean-reverting Gaussian process (Detemple and Osakwe, 2000), mean-reverting 3/2 volatility model (Goard and Mazur, 2013), 4/2 volatility model (Grasselli, 2017) and the generalized mixture model (Detemple and Kitapbayev, 2018). 
From a statistical/econometric perspective, several studies have indicated that the estimated value of the power of deviation is close to 1.5 
(Bakshi et al., 1997; Chacko and Viceira, 2003; Jones, 2003; Ishida and Engle, 2002; Goard and Mazur, 2013). These results imply that deviations from the standard 3/2 model are meaningful for capturing the complex aspects of VIX behavior.
Therefore, we model the volatility as a mean-reverting 3/2 volatility process in this paper.
 Goard and Mazur (2013) provided a closed-form pricing formula for the European volatility option in the mean-reverting 3/2 volatility process. 
The mean-reverting 3/2 volatility process is the reversed process of the MRSRP.
 Contrary to the MRSRP, Goard and Mazur (2013) also indicated that the mean-reverting 3/2 volatility process has a nonlinear drift so that it exhibits substantial nonlinear mean-reverting behavior when the volatility is above its long-run mean. Hence, after a large volatility spike, the volatility can potentially quickly decrease, while after a low volatility period it can be slow to increase.

With COVID-19, the VIX  jumped 50\% within a month and reached a record 82.69 on March 16, 2020 but was back down to 33.29 by April 27, 2020.
However, the VIX surged 12\% after President Trump tested positive for COVID-19.
This means that the buyer and the seller must also face the huge uncertainty due to the different types of volatility in turbulent times. 
The right to exercise prior the maturity can help the buyer to reduce possible volatility risk and increase the profit from American volatility options.
Recently, properties of the American volatility option have been  considered by Detemple and Osakwe (2000), Detemple and Kitapbayev (2018) and Liu (2015).
For the seller, an American option embedded with the callable feature will help the seller to reduce the risk arising from a dramatic increase in volatility and to decrease the hedging cost.
When the seller expects the volatility to rise, recalling the option might provide an efficient way to mitigate the undesirable position.
Moreover, the volatility option can not be hedged by holding a delta-hedge amount of the volatility index since the volatility index is not a tradable asset. 
Therefore, the callable American volatility option provides the seller with a useful instrument to construct a super-hedging strategy.

\subsection{Setup}
The $\delta$-penalty callable American volatility option offers the seller an embedded costly cancellation option which permits the seller to pay a penalty $\delta$ in addition to the payoff at the time of the cancellation. In this paper, we investigate the valuation problem for the $\delta$-penalty callable American volatility option in the mean-reverting 3/2-volatility processes.
The mean-reverting 3/2-volatility process under the martingale measure $Q$ is given as follows:
$$
dx_t=(\alpha x_t-\beta x_t^2) dt+k x_t^{3/2} dz
$$
where $\alpha$, $\beta$, $k>0$ are constants with $\alpha>r$ and $dz$ denotes the increment in the Wiener process under the martingale measure $Q$.
The coefficient $\beta$ is positive and the volatility index $x$ always remains positive (Goard and Mazur, 2013).

 Let $u(x_t)$ denote the value of 
a perpetual callable American volatility put with a cancellation feature available to the short side of the contract with penalty $\delta$. The payoff to the holder upon cancellation is
$(K-x)^++\delta$ when $x_t=x$. 
Let $g_1(x)=(K-x)^+$ and $g_2(x)=(K-x)^+ + \delta$ be two continuous functions with $0\leq g_1(x) < g_2(x)$. A continuous time Dynkin game with two players, a buyer and a seller, is described as follows: the buyer chooses a stopping time $\tau$ and the seller chooses a stopping time $\gamma$. At the time $\tau \wedge \gamma:= \min \{\tau, \gamma \}$, the seller pays the amount
$$
g_1(x_\tau)\cdot 1_{\tau< \gamma}+g_2(x_\gamma)\cdot 1_{\gamma<\tau}
$$
to the buyer. Then the value $u$ of this game is defined as
$$
u(x)=\sup_{\tau} \inf_{\gamma} \mathrm{E}_x R(\tau, \gamma)
$$
where $R(\tau,\gamma)=e^{-r\tau}g_1(x_\tau)\cdot 1_{\tau< \gamma}+e^{-r\gamma}g_2(x_\gamma)\cdot 1_{\gamma<\tau}$.
The value $u$ satisfies
$$
g_1(x)\leq u(x) \leq g_2(x)
$$
and is continuous (Ekstr\"{o}m, 2006). Moreover, given $\tau^*=\inf\{t|u(x_t)=g_1(x_t)\}$ and $\gamma^*=\inf\{t|u(x_t)=g_2(x_t)\}$, the value satisfies 
$$
u(x)=\sup_{\tau} \mathrm{E}_xR(\tau,\gamma^*)=\inf_{\gamma}\mathrm{E}_xR(\tau^*,\gamma)=\mathrm{E}_xR(\tau^*,\gamma^*).
$$

The infinitesimal generator of the process $(e^{-rt}x_t)_{0<t<\infty}$ is given by
$$
\mathcal{L}=\frac{1}{2}k^2 x^3 \frac{d^2}{dx^2}+(\alpha x-\beta x^2) \frac{d}{d x}-r.
$$
Let $\Phi_1(x)=\Phi(\frac{r}{\alpha},A;\frac{B}{x})$ and $\Phi_2(x)=(\frac{B}{x})^{1-A}\Phi(\frac{r}{\alpha}+1-A,2-A;\frac{B}{x})$ be two independent solutions to the differential equation $\mathcal{L} f=0$,
where $A=2(1-\frac{\beta}{k^2})> 0$ and $B=\frac{2\alpha}{k^2}> 0$. Then the solution $f$ is expressed as
$$
f(x)=C_1\Phi_1(x)+C_2\Phi_2(x).
$$ 
The coefficients $C_1$ and $C_2$ are determined by the boundary conditions.
Here, the function $\Phi$ denotes the confluent hypergeometric function and is defined as
$$
\Phi(a,b;x)=\frac{\Gamma(b)}{\Gamma(a)\Gamma(b-a)}\int_0^1 e^{ux}u^{a-1}(1-u)^{b-a-1} du.
$$  
Since the first derivative of $\Phi$ is given as
$$
\frac{d}{dx}\Phi(a,b;x)=\frac{a}{b}\Phi(a+1,b+1;x),
$$
the first derivatives of the independent solutions $\Phi_1(x)$ and $\Phi_2(x)$ are obtained as follows
$$
\frac{d}{dx}\Phi(\frac{r}{\alpha},A;\frac{B}{x})=\frac{-rB}{\alpha A} x^{-2} \Phi(\frac{r}{\alpha}+1,A+1;\frac{B}{x})
$$
and
$$
\begin{array}{l}
\frac{d}{dx}\left[ (\frac{B}{x})^{1-A}\Phi(\frac{r}{\alpha}+1-A,2-A;\frac{B}{x}\right]\\
=B^{1-A} x^{A-2}
\left[
\begin{array}{l}
(A-1)\Phi(\frac{r}{\alpha}+1-A,2-A;\frac{B}{x})\\
-\frac{(\frac{r}{\alpha}+1-A)B}{(2-A)x}\Phi((\frac{r}{\alpha}+2-A,3-A;\frac{B}{x}))
\end{array}
\right].
\end{array}
$$

\subsection{Main contributions}
Let $v$ be the value of the perpetual American knock-out volatility put with barrier $K$ and rebate $\delta$.
If $\beta<\frac{1}{2}k^2$, the closed-form pricing formula $v$ is considered by the two cases, $x\in (s,K)$ and  $x\in (K,\infty)$.
In the case where $x\in (s,K)$, the value $v$ is given as
$$
v(x)=
\left\lbrace
\begin{array}{ll}
K-x,& 0<x<s, \\
C_1\Phi_1(x)+C_2\Phi_2(x),& s<x<K,
\end{array}
\right.
$$
where $C_1$ and $C_2$ are given as follows
$$
C_1=\frac{(K-s)\Phi_2(K)-\delta \Phi_2(s)}{\Phi_1(K)\Phi_2(s)-\Phi_1(s)\Phi_2(K)},
$$
$$
C_2=\frac{\delta \Phi_1(s)-(K-s)\Phi_1(K)}{\Phi_1(K)\Phi_2(s)-\Phi_1(s)\Phi_2(K)}.
$$
The early exercise boundary  $s$ is a root of the following equation
$$
\begin{array}{l}
C_1\frac{-rB}{\alpha A} s^{-2} \Phi(\frac{r}{\alpha}+1,A+1;\frac{B}{s})\\
+C_2 B^{1-A} s^{A-2}
\left[
\begin{array}{l}
(A-1)\Phi(\frac{r}{\alpha}+1-A,2-A;\frac{B}{s}) \\
-\frac{(\frac{r}{\alpha}+1-A)B}{2-A} s^{-1} \Phi(\frac{r}{\alpha}+2-A,3-A;\frac{B}{s})
\end{array}
\right]
=-1
\end{array}
$$
In the case where $x\in (K,\infty)$,  the perpetual American knock-out volatility put will not be exercised early. The value $v$ is then given as
$$
v(x)=\delta\frac{\Phi_2(x)}{\Phi_2(K)}.
$$ 

Let $u$ be the value of the perpetual callable American volatility put. Define 
$$
\delta^*=v_A(K)\ \mathrm{and}\ a=v'(K),
$$
where $v_A$ is the value of the perpetual American volatility put.
If $\beta<\frac{1}{2}k^2$, the value $u$ of the perpetual callable American volatility put is considered by the following three cases:
\begin{enumerate}
\item If $\delta\geq \delta^*$, the $u(x)=v_A(x)$, for all $x\in (s,\infty)$.
\item If $\delta<\delta^*$ and $a\geq -1$, then $u(x)=v(x)$, for all $x\in [0,K]$.
\item If $\delta<\delta^*$, then $u(x)=v(x)$ for all $x\in [K,\infty]$.
\end{enumerate}
The proofs and the numerical results are demonstrated in the following sections.
\section{Proofs of our contributions}
In the first three subsections, we first review the pricing problem for the perpetual American volatility put and then 
provide the pricing formula for the perpetual American knock-out put option. 
In Section 2.4, we show that the value of the perpetual callable American put is equal to the value of the perpetual American knock-out put in the given conditions. 
The numerical results are demonstrated in Section 2.5.
\subsection{The perpetual American volatility put}
Let
$$
v_A(x)=\sup_\tau E_x[e^{-r \tau} g_1(x_{\tau})].
$$
be the value of the perpetual American volatility put. Liu (2015) provided a closed-form pricing formula for the perpetual American volatility put
by solving the following free boundary problem 
$$
\mathcal{L}v_A=0,\ \ \  x\in (s,\infty)
$$
with $v_A(s)=K-s$, $v_A'(s)=-1$ and $\lim_{x \rightarrow \infty} v_A(x)=0$.
The solution is given as
$$
v_A(x)=
\left\lbrace
\begin{array}{ll}
K-x,& x<s, \\
\frac{\Phi_2(x)}{\Phi_2(s)}(K-s)(\frac{s}{x})^{A-1},& s<x<\infty,
\end{array}
\right.
$$
where $s$ satisfies the following equation
$$
-\frac{\Phi(\frac{r}{\alpha}+1-A,2-A;\frac{B}{s})}{s^{-1}\left[
\begin{array}{l}
(A-1)\Phi(\frac{r}{\alpha}+1-A,2-A;\frac{B}{s})\\
-\frac{(\frac{r}{\alpha}+1-A)B}{(2-A)s}\Phi((\frac{r}{\alpha}+2-A,3-A;\frac{B}{s}))
\end{array}
\right]}=K-s.
$$

In addition, Detemple and Kitapbayev  (2018) considered the American volatility put in the mixture volatility process, where the mixture process adds the mean-reverting square root process and the mean-reverting 3/2 volatility process.
They show that the early exercise price satisfies the system of integral equations.
  
\subsection{The perpetual knock-out American volatility put}
An American knock-out option is an American option for which the holder receives a certain amount of money when the knock out occurs.
For this option, the holder can seek an optimal strategy to maximize the option's value before the volatility index reaches a certain price.
Define 
$$
v(x)=\sup_\tau E_x[e^{-r\tau} (g_1(x_\tau)\cdot 1_{\tau< \tau_K}+\delta \cdot 1_{\tau=\tau_K})]
$$
where $\tau_K=\inf\{t|x_t=K\}$. 
The function $v$ is regarded as the value of a perpetual knock-out put with barrier $K$ and rebate $\delta$. 
Then $v$ as well as the early exercise boundary $s$ satisfies the following free boundary problem
$$
\mathcal{L} v=0,\ \ \  s<x<K
$$
with $v(s)=K-s$, $v(K)=\delta$ and $v'(s)=-1$.
To provide the pricing formula, we first consider the case where $x<K$.
\begin{theorem}
Let $v(x)$ be the value of a perpetual knock-out American volatility put with barrier $K$ and rebate $\delta$. If $\beta<\frac{1}{2}k^2$, the closed-form pricing formula is obtained as
\begin{equation} \label{eq:KO}
v(x)=
\left\lbrace
\begin{array}{ll}
K-x,& 0<x<s, \\
C_1\Phi_1(x)+C_2\Phi_2(x),& s<x<K,
\end{array}
\right.
\end{equation}
where $C_1$ and $C_2$ are given as follows
$$
C_1=\frac{(K-s)\Phi_2(K)-\delta \Phi_2(s)}{\Phi_1(K)\Phi_2(s)-\Phi_1(s)\Phi_2(K)},
$$
$$
C_2=\frac{\delta \Phi_1(s)-(K-s)\Phi_1(K)}{\Phi_1(K)\Phi_2(s)-\Phi_1(s)\Phi_2(K)}.
$$
The early exercise boundary  $s$ is the root of the following equation
\begin{equation} \label{eq: equ for early price Knock}
\begin{array}{l}
C_1\frac{-rB}{\alpha A} s^{-2} \Phi(\frac{r}{\alpha}+1,A+1;\frac{B}{s})\\
+C_2 B^{1-A} s^{A-2}
\left[
\begin{array}{l}
(A-1)\Phi(\frac{r}{\alpha}+1-A,2-A;\frac{B}{s}) \\
-\frac{(\frac{r}{\alpha}+1-A)B}{2-A} s^{-1} \Phi(\frac{r}{\alpha}+2-A,3-A;\frac{B}{s})
\end{array}
\right]
=-1
\end{array}
\end{equation}
\end{theorem}
\begin{proof}
If $\beta<\frac{1}{2}k^2$, the general solution to $\mathcal{L}v=0$ is written by using the two independent confluent hypergeometric functions as follows
$$
v(x)=C_1\Phi_1(x)+C_2\Phi_2(x)
$$
where $s<x<K$. Substituting the boundary conditions into the general solution yields the following linear system
$$
C_1\Phi_1(s)+C_2\Phi_2(s)=K-s
$$
and
$$
C_1\Phi_1(K)+C_2\Phi_2(K)=\delta.
$$
Solving the linear system, the values of $C_1$ and $C_2$ can be determined in terms of $\Phi_1(s)$, $\Phi_2(s)$, $K$, $s$ and $\delta$. 
Using the high-contact condition $v'(s)=-1$, the early exercise price $s$ is a root of (\ref{eq: equ for early price Knock}).
\end{proof}
Since the value $v$ decreases to zero as $x$ tends to infinity, the buyer does not exercise the option prior to its maturity in the case where $x\geq K$.
Hence, the value of the perpetual American knock-out volatility put satisfies the
following equation
\begin{equation}\label{eq: x>K}
\mathcal{L}v=0,\ \ \ K<x<\infty,
\end{equation}
with $v(K)=\delta$ and $\lim_{x\rightarrow \infty}v(x)=0$.
The value of a perpetual knock-out volatility put is obtained in the following theorem. 
\begin{theorem}
Let $v$ be the solution of (\ref{eq: x>K}). If $\beta<\frac{1}{2}k^2$, then $v$ is given as
$$
v(x)=\delta\frac{\Phi_2(x)}{\Phi_2(K)}.
$$ 
\end{theorem}
\begin{proof}
If $\beta<\frac{1}{2}k^2$, the solution of $\mathcal{L}v=0$ is expressed as 
$$
v(x)=C_1\Phi_1(x)+C_2\Phi(x).
$$
Since $\Phi(\frac{r}{\alpha},A;B/x)\neq 0$ as $x\rightarrow \infty$, the boundary condition $\lim_{x\rightarrow \infty} v(x)=0$ forces $C_1=0$.
Hence, we have $v(x)=C_2\Phi_2(x)$. Substituting $v(K)=\delta$ into $v(x)=C_2\Phi_2(x)$ yields 
$$
C_2=\frac{\delta}{\Phi_2(K)}.
$$
This implies that $v(x)=\delta\frac{\Phi_2(x)}{\Phi_2(K)}$ for the case where $x> K$.
\end{proof}
\subsection{Properties of the perpetual American knock-out volatility put}
In this section, we consider the properties for $v(x)\leq g_2(x)$ on $(s,\infty)$. 
We first consider the case where $x\in (K,\infty)$.
\begin{theorem}
Let $v$ be a solution of $\mathcal{L}v=0$, $x\in (K,\infty)$ with $v(K)=\delta$ and $\lim_{x\rightarrow \infty} v(x)=0$. Then we have $v(x)<\delta$ for all $x\in (K,\infty)$.
\end{theorem}
\begin{proof}
By letting $f(x)=u-\delta$ on $(K,\infty)$, then $f(K)=0$ and $\lim_{x\rightarrow \infty}f(x)<0$. 
Applying the operator $\mathcal{L}$ to $f$ yields
$$
\mathcal{L} f = -r\delta,\ \ \forall x\in (K,\infty).
$$
This implies that $f(x)$ has no nonnegative local maximum at $c\in (K,\infty)$. Hence, we have $u(x)-\delta<0$ for all $x\in (K,\infty)$.
That is, $u(x)\leq \delta$ on $[K,\infty)$.
\end{proof}

For the case where $x\in (s,K)$, we obtain two theorems that make $v(x)\leq g_2(x)$.
Applying $\mathcal{L}$ to $g_2(x)$, we obtain
$$
\mathcal{L} g_2(x)=
\left\lbrace
\begin{array}{ll}
\beta x^2-(\alpha-r)x-r(K+\delta)& x<K, \\
-r\delta& x>K.
\end{array}
\right.
$$
The roots of the first quadratic equation can be expressed as
$$
\frac{(\alpha-r)\pm\sqrt{(\alpha^2+4r(K+\delta)\beta)}}{2\beta}.
$$
Since $\alpha>r$, the solution 
$$
\frac{(\alpha-r)-\sqrt{(\alpha^2+4r(K+\delta)\beta)}}{2\beta}<0
$$
does not belong to $[0,\infty)$. Thus, the desired root is
$$
d_1=\frac{(\alpha-r)+\sqrt{(\alpha^2+4r(K+\delta)\beta)}}{2\beta}>0.
$$
Hence, we have
$$
\beta x^2-(\alpha-r)x-r(K+\delta)\leq 0, \forall x\in (0,d_1].
$$ 

\begin{theorem}\label{THM: x<K}
Let $v$ be a solution of the following differential equation
$$
\mathcal{L}v=0,\ \ x\in (s,K)
$$
with $v(K)=\delta$ and $v(s)=K-s$.
If $K \leq d_1$, then $v(x)\leq K-x+\delta=g_2(x) $ for all $x\in (s,K)$. 
\end{theorem}
\begin{proof}
By letting $f(x)=v(x)-g_2(x)$ on $[s,K]$, we have then $f(s)=-\delta$ and $f(K)=0$.
To show $v(x)\leq K-x+\delta$ on $(s,K)$, we apply the operator $\mathcal{L}$ to $f$ and obtain
$$
\mathcal{L}f=-(\beta x^2-(\alpha-r)x-r(K+\delta)), \forall s< x< K.
$$
Since $K\leq d_1$, we have $\mathcal{L}f> 0$. 
By the maximum principle, we have $u(x)\leq K-x+\delta =g_2(x)$ on $(s,K)$ if $K<d_1$.
\end{proof}

\begin{theorem}\label{THM: x<K 2}
Let $v$ be the value of the perpetual American knock-out put with the barrier $K$ and the rebate $\delta$.
If $v'(K)=a>-1$ and $2\beta x+r-\alpha>0$ on $(s,K)$, then $-1<v'(x)<a$ on $(s,K)$. Moreover, we have $v(x)\leq g_2(x)$ on $(s,K)$.
\end{theorem}
\begin{proof}
Let $w=v'$. Since $\mathcal{L}v=0$ on $(s,K)$, we have
$$
0=\frac{d}{dx}\mathcal{L}v=\frac{1}{2}k^2 x^3 \frac{d^2 w}{dx^2}+(\frac{3}{2}k^2 x^3 -\beta x^2 +\alpha x) \frac{d w}{d x}-(2\beta x+r-\alpha)w.
$$
The boundaries of $w$ are obtained as $w(s)=v'(s)=-1$ and $w(K)=v'(K)=a>-1$. By the maximum principle, we have 
$-1<w(x)<a$ on $(s,K)$.

Moreover, if there exists a $b\in (s,K)$ such that $v(b)=m>g_2(x)$, then the value of $v$ rises to touch $m>K-b+\delta$ and falls down to $\delta$.
There must exist a $c\in (s,K)$ such that $v'(c)<-1$, since $v(K)=\delta=g_2(K)$ and $v(s)=K-s<K-s+\delta=g_2(s)$. This contradicts $-1<v'(x)<a$ on $(s,K)$.
Therefore, we have $v(x)\leq g_2(x)$ on $(s,K)$.
\end{proof}

\subsection{The perpetual callable American volatility put}
A callable American volatility option is a contract that gives the buyer and the seller the right to 
seek an optimal strategy to maximize (for the buyer) and to minimize (for the seller) the option's value.
Hence the value of the perpetual callable American volatility option satisfies 
$$
u(x)=\sup_{\tau} \inf_{\gamma} \mathrm{E}_x R(\tau, \gamma)
$$
where $R(\tau,\gamma)=e^{-r\tau}g_1(x_\tau)\cdot 1_{\tau< \gamma}+e^{-r\gamma}g_2(x_\gamma)\cdot 1_{\gamma<\tau}$.

Let $v$ be the value of the perpetual American knock-out volatility put with barrier $K$ and rebate $\delta$.
Define 
$$
\delta^*=v_A(K),
$$
and 
$$
a=v'(K),
$$
where $v_A$ is the value of the perpetual American volatility put.
Then, we obtain the following result.

\begin{theorem}\label{THM: main}
Let $u$ be the value of the perpetual American volatility put. We consider the following three cases:
\begin{enumerate}
\item If $\delta\geq \delta^*$, the $u(x)=v_A(x)$, for all $x\in (s,\infty)$.
\item If $\delta<\delta^*$ and Theorem \ref{THM: x<K} (or Theorem \ref{THM: x<K 2}) holds, then $u(x)=v(x)$, for all $x\in [0,K]$.
\item If $\delta<\delta^*$, then $u(x)=v(x)$ for all $x\in [K,\infty]$.
\end{enumerate}
\end{theorem}
\begin{proof}
The proof of Case 1 is the same as that for the callable American put.
For Case 2, we have that $\delta<\delta^*$ and $v(x)<g_2(x)$ in $(s,K)$. 
To show that $u=v$ on $(s,K)$, it suffices to show that $K$ is the optimal strategy minimizing the option's value. 
Suppose that the seller calls back (or cancels) the option at some particular price $b<K$. 
Then, we have $u(x)=K-x+\delta$ for all $x\in (b,K)$. 
However, we have $v(x)\leq K-x+\delta=u(x)$ if $K<d_1$ for Theorem \ref{THM: x<K} (or if $a>-1$ for Theorem \ref{THM: x<K 2}). 
This implies that cancelling the option at $K$ means that a lower value is obtained than cancelling at $b$. 
Therefore, cancelling the option at any price $b<K$ is not the optimal strategy to minimize the option's value under the given conditions.
 As a result, the value of a perpetual callable American volatility put is equal to the value of a perpetual knock-out American volatility put if $K<d_1$. 
Based on a similar argument, we obtain that $u(x)=v(x)$ for all $x\in (K,\infty)$ for Case 3.
\end{proof}

\subsection{Numerical results}
\begin{figure}[h]
  \centering
  \includegraphics[width=80mm]{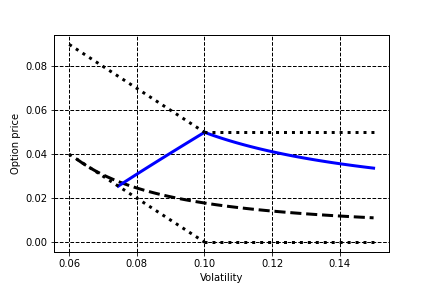}
  \caption{Comparison of the values between the perpetual callable American volatility put $v$(solid line) and the perpetual knock-out American volatility put $u=v_A$(dashed line) with $\delta=0.05$}
  \label{fig:1}
\end{figure}

Let the values for the volatility of volatility $k$, interest rate $r$ and strike $K$ be given as  $\beta=0.2$, $r=0.05$, and $K=0.5$.
The graph of the case where $v_A(K)>\delta$ (with $\alpha=0.001$, $\sigma=0.5$, $\delta=0.05$) is presented in Fig. \ref{fig:1}. 
We find that the value $v$ of the perpetual American knock-out volatility put
increases to the rebate $\delta$ in the neighborhood of $K$ when the rebate $\delta$ is greater than $v_A(K)$. Hence, the first derivative of $v$ is positive near $K$ as $x<K$. 
However, the value $v_A$ of the perpetual American volatility put decreases from $v_A(s)=K-s$ to zero as $x$ tends to infinity and $v_A(K)<\delta=v(K)$.
Therefore, the rational seller may not cancel the American volatility put by paying the penalty $\delta>v_A(K)$.
This implies that we have
$u=v_A$ for the case where $v_A(K)>\delta$ just as in Theorem \ref{THM: main}, where $u$ denotes the value of the perpetual callable American volatility put.  

\begin{figure}[h]
  \centering
  \includegraphics[width=80mm]{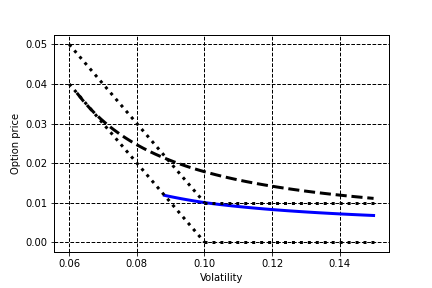}
  \caption{Comparison of the values between the perpetual callable American volatility put $v$(solid line) and the perpetual knock-out American volatility put $u=v_A$(dashed line) with $\delta=0.05$}
  \label{fig:3}
\end{figure}

Based on the same values for $\alpha$, $\sigma$, $\beta$, $r$ and $K$, we consider the case where $v_A(K)<\delta$ and $v'(K)=a>-1$ on $(s,K)$ with $\delta=0.01$.
The graph is demonstrated in Fig. \ref{fig:3}. The value $v_A$ of the perpetual American volatility put crosses over $g_2(x)$ at $b\in (s,K)$ and is greater than $g_2(x)$ on $(b,K)$. The value $v$ of the perpetual American knock-out volatility put is less than $g_2(x)$ for all $x\in (0,\infty)$. 
Therefore, the rational seller will cancel the American volatility put by paying $\delta<v_A(K)$. 
This implies that we have $u(x)=v(x)$ on $(s,\infty)$ just as in Theorem \ref{THM: main} for the case where $v_A(K)<\delta$ and $v'(K)=a>-1$ on $(s,K)$.

\section{Conclusion}
Volatility derivatives are an important innovation in the field of finance.
For participants in the market, this paper introduces a callable American volatility put to reduce the volatility risk in turbulent times.  
Moreover, the callable feature decreases the hedging cost for the seller and increases the attractiveness for the buyer.
An explicit valuation formula is derived for the case of the perpetual knock-out American volatility put and perpetual callable American volatility put.
The numerical results demonstrate that the value of the perpetual callable American volatility put is lower than the value of the corresponding perpetual American volatility option.

\appendix


\end{document}